\documentclass[a4paper]{aa}
\usepackage{graphicx,float,multirow}

\def \src {IGR\thinspace{J17544$-$2619}}

\def \nh {$N{\rm _H}$}

\def \hcm {\hbox {\ifmmode $ atom cm$^{-2}\else atom cm$^{-2}$\fi}}
\def \arcmin {\hbox{$^\prime$}}
\def \arcsec {\hbox{$^{\prime\prime}$}}
\def \chisq {$\chi ^{2}$}

\def\approxgt{\mathrel{\hbox{\rlap{\lower.55ex \hbox {$\sim$}}
        \kern-.3em \raise.4ex \hbox{$>$}}}}
\def\approxlt{\mathrel{\hbox{\rlap{\lower.55ex \hbox {$\sim$}}
        \kern-.3em \raise.4ex \hbox{$<$}}}}
\newcommand{\mc}{\multicolumn}

\begin{document}

\title{XMM-Newton observations of the INTEGRAL X-ray transient 
\src}

\author{R. Gonz\'alez-Riestra\inst{1} 
    \and T. Oosterbroek\inst{2}
    \and E. Kuulkers\inst{2}
    \and A. Orr\inst{2}
    \and A. N. Parmar\inst{3}}

\offprints{R. Gonz\'alez-Riestra,
 \email{rgonzale@xmm.vilspa.esa.es}}

\institute{XMM-Newton Science Operation Centre, ESA, VILSPA,
          P.O.Box 50727, E-28080 Madrid, Spain
\and
       INTEGRAL Science Operations Centre,
       Science Operations and Data Systems Division,
       Research and Scientific
       Support Department of ESA, ESTEC,
       Postbus 299, NL-2200 AG Noordwijk, The Netherlands
\and
       Astrophysics Missions Division, Research and Scientific
       Support Department of ESA, ESTEC,
       Postbus 299, NL-2200 AG Noordwijk, The Netherlands
}

\date{Received ; Accepted:}

\authorrunning{R. Gonzalez-Riestra et al.}

\titlerunning{XMM-Newton observations of \src}

\abstract{On 2003 September 17 INTEGRAL discovered a bright
transient source 3$\degr$ from the Galactic Center, \src. The field
containing the transient was observed by XMM-Newton on 2003 March 17
and September 11 and 17.  A bright source, at a position consistent
with the INTEGRAL location, was detected by the European Photon
Imaging Camera (EPIC) during both September observations with mean
0.5--10\,keV unabsorbed luminosities of 1.1$\times$10$^{35}$ and
5.7$\times$10$^{35}$\,erg\,s$^{-1}$ for an (assumed) distance of
8\,kpc.  The source was not detected in 2003 March with a 0.5--10\,keV
luminosity of $<$3.8$\times$10$^{32}$\,erg\,s$^{-1}$.  The September
11 and 17 EPIC spectra can be represented by a power-law model with
photon indices of 2.25$\pm$0.15 and 1.42$\pm$0.17, respectively. Thus,
the 0.5--10\,keV spectrum hardens with increasing intensity. The
low-energy absorption during both September observations is comparable
to the interstellar value.  The X-ray lightcurves for both September
observations show energy dependent flaring which may be modeled by
changes in either low-energy absorption {\it or} power-law index.
}
\maketitle

\section{Introduction}
\label{sect:intro}

About a dozen new hard X-ray transients have been discovered in the
last year during INTEGRAL (Winkler et al.~\cite{w:03}) observations of
the galactic center region with the soft gamma-ray imager IBIS/ISGRI
(Lebrun et al.~\cite{le:03}).  Their unusual spectral hardness has led
to suggestions that these sources comprise a group of highly absorbed
galactic binaries (Revnivtsev et al.~\cite{re:03}). These are being
preferentially detected due to the good sensitivity and large field of
view of IBIS/ISGRI above 15\,keV. XMM-Newton observations of the first
of these, IGR\,J16318$-$4848, revealed intense Fe~K$\alpha$ and
K$\beta$ and Ni~K$\alpha$ emission lines as well as strong low-energy
absorption (Matt \& Guainazzi~\cite{mg:03}; Walter et
al.~\cite{wa:03}).  XMM-Newton observations of IGR\,J16320$-$4851
revealed a featureless hard continuum (Rodriguez et
al.~\cite{r:03b}). IGR\,16358$-$4726 was observed by {\it Chandra} and
showed a hard power-law spectrum with a 5880\,s periodic intensity
modulation (Patel et al.~\cite{p:03}).

\begin{table*}

\caption[]{XMM-Newton observation log. The EPIC modes are Full Frame
(FF) and Timing (TI).  In the 2003 March 17 observation \src\ was
outside the OM field of view. The effective wavelengths of the UVW1,
UVW2 and UVM2 filters are 2945\,\AA, 2180\,\AA, and 2340\,\AA,
respectively.  }
\begin{flushleft}
\begin{tabular}{lccccccccc}
\hline
\hline\noalign{\smallskip}
Obs & Start Time & End Time & \mc{2}{c}{Exposure (ks)} & \mc{3}{c}{EPIC 
mode} & \mc{2}{c}{OM} \\
    &     (UTC)  &   (UTC)  & pn &MOS                  &  pn & MOS1 & 
MOS2    & Exp (ks) & Filter\\
\noalign{\smallskip\hrule\smallskip}
1 & 2003~Mar~17~20:53 & Mar 18~00:33 & 10.2 & 11.9 & FF  & FF & FF 
& \dots & \dots   \\
2 & 2003~Sep~11~17:54 & Sep~11~22:09 & 9.3  & 11.0 & FF  & FF 
& FF & 2.2, 3.2 & B, UVW1  \\
3 & 2003~Sep~17~17:13 & Sep~17~19:59 & 2.5  & 8.3  & TI  & TI 
    & FF & 1.8, 1.6 & UVM2, UVW2 \\
\noalign{\smallskip\hrule\smallskip}
\end{tabular}
\end{flushleft}
\label{tab:obs}

\end{table*}

We report on a new bright transient source, \src, discovered using
IBIS/ISGRI on 2003 September 17 at 01:10 UTC during an observation of
the Galactic Center region (Sunyaev et al.~\cite{s:03}).  The source
intensity was about 160\,mCrab, 60\,mCrab, and $<$15\,mCrab (at
3$\sigma$ confidence) in the 18--25\,keV, 25--50\,keV and 50--100\,keV
energy ranges. \src\ was bright for around 2 hours and then faded
below the IBIS/ISGRI detection threshold. The source was again
detected in outburst by IBIS/ISGRI later the same day between 06 and
14~hrs~UTC (Grebenev et al.~\cite{g:03}).

By chance, XMM-Newton observed the region of sky containing \src\ only
5 days before the INTEGRAL discovery. A source was clearly detected at
a position consistent with that reported from INTEGRAL, with a mean
2--10\,keV intensity of 4.5$\times$10$^{-12}$\,erg\,cm$^{-2}$~s$^{-1}$
(Gonz\'alez-Riestra et al.~\cite{gr:03}).  Rodriguez et
al.~(\cite{r:03a}) reported on a possible optical/infrared counterpart
in the USNO~B1.0 catalog (B = 13.9--14.5$\pm$0.3~mag) and 2MASS
all-sky quick-look image archive.  They also noted that there are 3
fainter candidates within the preliminary 10\arcsec\ XMM-Newton
uncertainty region in the 2MASS image.

The field containing \src\ was observed three times by XMM-Newton. The
first two observations (2003 March 17 and September 11) were part of a
program to study the nova V4643\,Sgr (see also Gonz\'alez-Riestra et
al.~\cite{gr:03}), and the third was a Target of Opportunity
observation triggered by the INTEGRAL discovery. Here, we present
results from all three XMM-Newton observations.

\begin{figure*}
\begin{center}
  \includegraphics[height=17cm,angle=90]{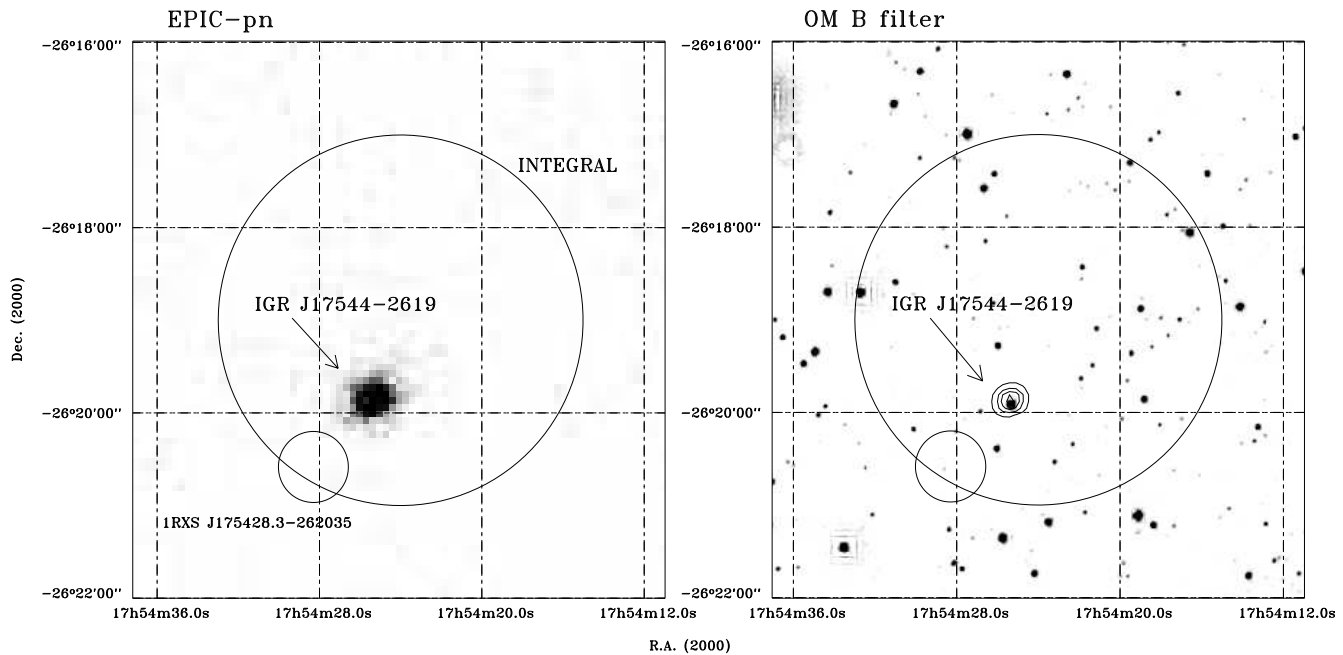}
  \caption[]{2003 September 11 EPIC-pn (left) and OM B filter (right)
  images of the area around the 2\arcmin\ INTEGRAL uncertainty region
  for IGR J17544-2619 of Sunyaev et al. (\cite{s:03}, big circle). The
  contours shown in the optical image correspond to the EPIC-pn image
  (0.005, 0.010, 0.015, 0.020 and 0.025 counts~s$^{-1}$), which
  clearly excludes 1RXS\,J175428.3-262035 (small circle).}
  \label{fig:EPIC_image}
\end{center}
\end{figure*}

The XMM-Newton Observatory (Jansen et al.~\cite{j:01}) includes three
1500\,cm$^2$ X-ray telescopes each with an European Photon Imaging
Camera (EPIC) at the focus. Reflection Grating Spectrometers (RGS, den
Herder et al.~\cite{dh:01}) are located behind two of the
telescopes. In addition, a coaligned optical/UV Monitor (OM, Mason et
al.~\cite{m:01}) is included.  Two of the EPIC imaging spectrometers
use MOS CCDs (Turner et al.~\cite{t:01}) and one uses a pn CCD
(Str\"uder et al.~\cite{s:01}).

\section{Observations and Data Reduction}
\label{sect:obs}

Table~\ref{tab:obs} gives observing times, exposures and instrument
modes.  In the first two observations the three EPIC detectors were
operated in the Full Frame mode, while during the third observation
the pn and the MOS1 were operated in Timing Mode. In this mode, data
are collapsed into a one dimensional row to be read at high speed with
the second dimension being replaced by timing information.  This
allows time resolutions of 1.5\,ms and 0.03\,ms for the MOS and pn,
respectively. Except for the 2003 March MOS2 observation, performed
with the medium filter, the EPIC thin filters were used throughout.

EPIC data were processed with the standard XMM-Newton SAS tasks
'epchain' and 'emchain'.  For the pn and MOS imaging data, source
counts were extracted from circular regions of 40\arcsec\ radius
centered on \src. Background counts were obtained from a similar
region offset from the source position for pn, and from an annulus
around the source for MOS2. Spectra were extracted separately for
single, double and single+double events.  For timing data, the source
spectra were extracted from columns centered on the source position,
and the background spectra from offset columns at both sides of the
source.

\begin{figure*}
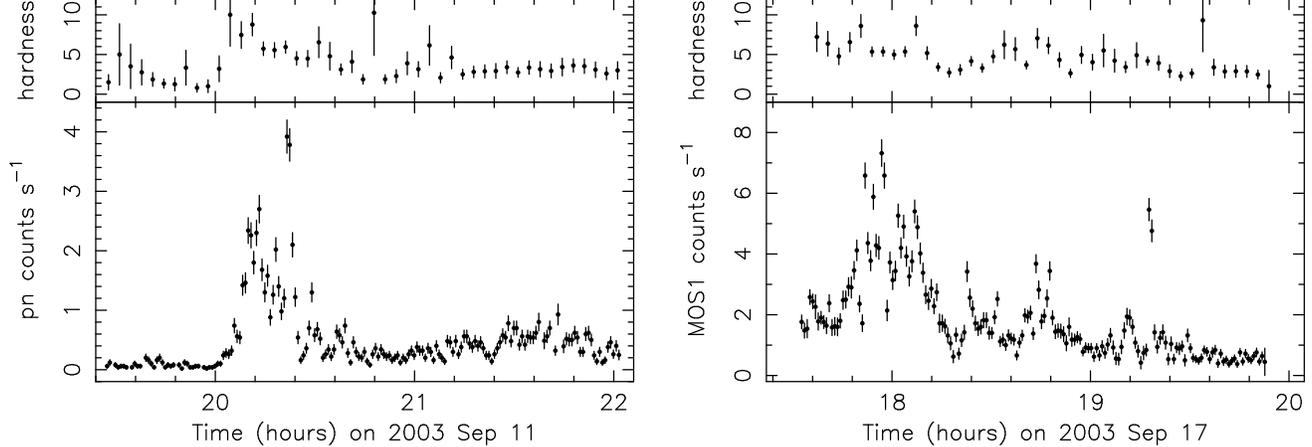

  \hbox{\hspace{0.4cm}
  \includegraphics[height=8.5cm,angle=-90,bb=186 54 578 593]{sep11_lc_hc.ps}
  \hspace{0.2cm}
  \includegraphics[height=8.5cm,angle=-90,bb=186 54 578 593]{sep17_lc_hc.ps}}
  \caption[]{\src\ 0.5--10\,keV light curves for the 2003 September
             11 (pn; lower left) and September 17 (MOS1; lower right) 
             observations with integration times of 50\,s. 
             A MOS1 count rate of 1 count s$^{-1}$
             corresponds to $\sim$3.5 pn count s$^{-1}$ (see text).
             The upper panels show the
             hardness ratio (counts between 2--10\,keV divided by
             those between 0.5--2\,keV) with a 200\,s binning.}
  \label{fig:lc}
\end{figure*}

The XMM-Newton OM (Optical Monitor) was operated in imaging mode
during the three observations (see also Table~1). During the 2003
March observation \src\ was outside the OM field of view. The OM data
were processed with the standard SAS task 'omichain'.

\section{Results}

\subsection{Source Position}

Unfortunately, during the 2003 September 11 observation \src\ was in
one of the gaps between the MOS CCDs. In the September 17
observation the MOS2 data suffer slightly from the effects of pile-up.
The pn data of 2003 September 11, therefore, provide the best estimate
of the (J2000) source position of R.A. =17$^{\rm h}$ 54$^{\rm m}$
25\fs37 Decl. =$-$26$^{\rm o}$ 19$'$ 52\farcs9, with a 90\%\
confidence uncertainty radius of 4\arcsec. This position improves on
that presented in Gonzalez-Riestra et al.\ (2003), which was based on
a preliminary estimate of the spacecraft attitude and clearly excludes
1RXS J175428.3-2620 (Wijnands 2003).  Fig.~\ref{fig:EPIC_image} shows
the EPIC-pn and OM B filter images with the 2\arcmin\ INTEGRAL
uncertainty region of Sunyaev et al. (2003) superposed.

Examination of the September 11 OM data reveals that there is a source
close to the edge of the refined EPIC pn uncertainty region (see
Fig.~1) with a B magnitude of 14.57$\pm$0.02. This is almost certainly
the USNO~B1.0 catalog source (2MASS\,J17542527$-$2619526) discussed in
Rodriguez et al.~(\cite{r:03a}).  The OM source is 3\arcsec\ from the
EPIC pn best source position. There is a similar offset (4\arcsec\/)
between the OM and the EPIC pn coordinates of V4643\,Sgr.  Once a
correction for this offset is made, the difference between the EPIC pn
coordinates and those of the optical star inside the error circle is
only 1\farcs2. The same target is present in the ultraviolet image
with a UVW1 magnitude of 16.13$\pm$0.04, and in the UVW2 image of
September 17 with UVW2=14.49$\pm$0.02. The source is not detected in
the UVM2 image.

In the 2003 March observation only X-ray upper limits to the intensity
of a source at the position of \src\ are obtained. The 0.5--10\,keV pn
3$\sigma$ upper limit of 0.008 count~s$^{-1}$ corresponds to a flux of
5$\times$10$^{-14}$\,erg\,cm$^{-2}$\,s$^{-1}$ for a power-law spectrum
with a photon index, $\alpha$, of 2.0 and an $N_{\rm H}$ of
1.4$\times$10$^{22}$\,atom\,cm$^{-2}$ (the interstellar value in the
direction of \src; Dickey \& Lockman~\cite{d:90}).

\begin{figure}
  \includegraphics[height=8.5cm,angle=-90]{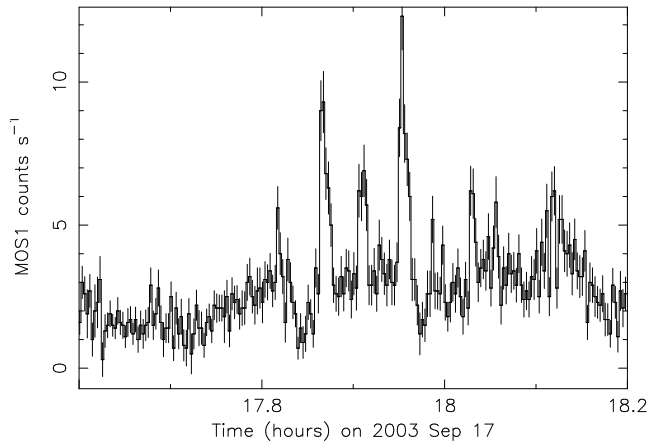}
  \caption[]{Enlarged view of the 2--10 keV September 17 light curve
  showing a number of short flares that took place between 17:45 and
  18:15 UTC (binning is 10 sec).}  
\label{fig:fl}
\end{figure}

\subsection{X-ray light curves}
\label{subsect:lightcurve}

The 0.5--10\,keV \src\ light curves and hardness ratios (counts in the
energy range 2--10\,keV divided by those between 0.5--2\,keV) from
both September observations are shown in Fig.~\ref{fig:lc}.  Since for
the September 17 observation pn data are available only after
19:15~UTC due to technical problems, we show here the MOS1 data.  In
order to compare the two light curves, it is worth noting that during
this overlapping interval 1 MOS1 count\,s$^{-1}$ corresponds to
$\sim$3.5 pn count\,s$^{-1}$.

During the September 11 observation the 0.5--10\,keV pn count
rate increased from 0.06\,s$^{-1}$ to $\sim$3\,s$^{-1}$ in about
500\,s.  A short ($\sim$100\,s) flare occurred at 20:25\,UTC when the
count rate reached its maximum of $\sim$4\,s$^{-1}$.  The count rate
then decreased by a factor $\sim$10 and remained around 0.4\,s$^{-1}$
for the rest of the observation. In general, changes in intensity were
accompanied by changes in the hardness ratio. \src\ was much brighter
during the September 17 observation with the underlying MOS1 count
rate decreasing from $\sim$2~s$^{-1}$ to $\sim$0.5~s$^{-1}$ with a
number of flares up to $\sim$7.5\,s$^{-1}$ superposed (see
Fig.~\ref{fig:fl}). The peak MOS1 count rate of $\sim$7.5\,s$^{-1}$
corresponds to $\sim$25~pn~s$^{-1}$, approximately a factor 6 higher
than during the  September 11 observation.

Power-density spectra of the September data reveal power-law shaped
noise components, consistent with flaring on various time scales, with
no obvious signatures such as pulsations or quasi-periodic
oscillations.

\subsection{X-ray Spectra}
\label{subsect:spectra}

All extracted spectra were rebinned to oversample the full width half
maximum of the energy resolution by a factor 3 and to have
additionally a minimum of 20 counts per bin to allow use of the
$\chi^2$ statistic. In order to account for calibration uncertainties
a 2\% error was added quadratically to each spectral bin. Spectral
uncertainties are given at 90\% confidence.

Initially, simple absorbed spectral models were fit to the
0.5--10\,keV September pn spectra integrated over each
observation. For the September 11 spectrum, a power-law with $\alpha =
2.25 \pm 0.15$ and $N _{\rm H} = (3.8 \pm 0.4) \times
10^{22}$\,atom\,cm$^{-2}$ gives a $\chi ^2$ of 151.9 for 142 degrees
of freedom (d.o.f.), a thermal bremsstrahlung with $kT = 5.5 \pm
0.9$\,keV and $N _{\rm H} = (3.20 \pm 0.25) \times
10^{22}$\,atom\,cm$^{-2}$ gives a $\chi ^2$/d.o.f.\ of 143.1/142 and
blackbody with $kT = 1.17 \pm 0.05$\,keV and $N _{\rm H} = (1.48 \pm
0.20) \times 10^{22}$\,atom\,cm$^{-2}$ gives a $\chi ^2$/d.o.f.\ of
141.0/142. Since all the models describe the spectra with comparable
goodness of fit, we cannot reliably distinguish between these simple
spectral models. We therefore use the results from the power-law fit
to estimate a mean 0.5--10\,keV unabsorbed flux of
1.5$\times$10$^{-11}$\,erg\,cm$^{-2}$\,s$^{-1}$. The September 17
spectrum can be represented by a power-law model with $\alpha = 1.42
\pm 0.09$ and $N _{\rm H} = (2.50 \pm 0.17) \times
10^{22}$\,atom\,cm$^{-2}$ to give a $\chi ^2$/d.o.f.\ of
283.8/215. The mean 0.5--10\,keV unabsorbed flux is
7.5$\times$10$^{-11}$\,erg\,cm$^{-2}$\,s$^{-1}$.  Examination of the
residuals indicates that the spectrum is not well fit
$\approxlt$1\,keV, which may indicate that there were variations in
\nh.  This spectrum is significantly harder than that of 
September 11.

\begin{figure}
  \includegraphics[height=8.5cm,angle=-90,bb=107 20 576 706]{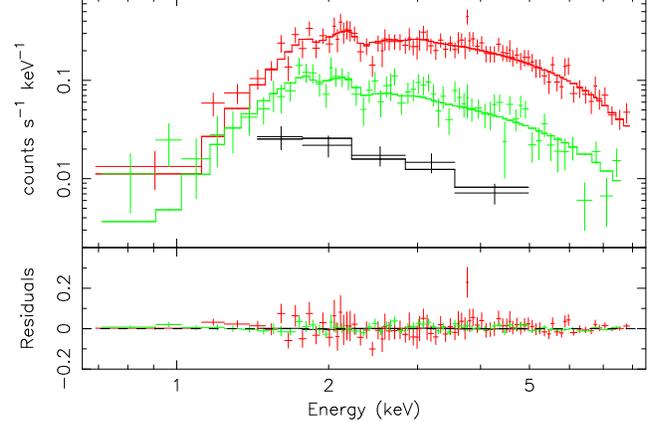}
  \caption[]{Time selected (see text) pn count spectra from the
             2003 September 11 observation. From bottom to top
             the 3 spectra correspond to pre-flare, post-flare and flaring
             intervals. The best-fit absorbed power-law models with the
             same value of $\alpha$ and different values of the
             normalisation and 
             \nh\ for each spectrum are shown as histograms.}
  \label{fig:sep11_spec}
\end{figure}

\begin{figure}
  \includegraphics[height=8.5cm,angle=-90,bb=107 20 576 706]{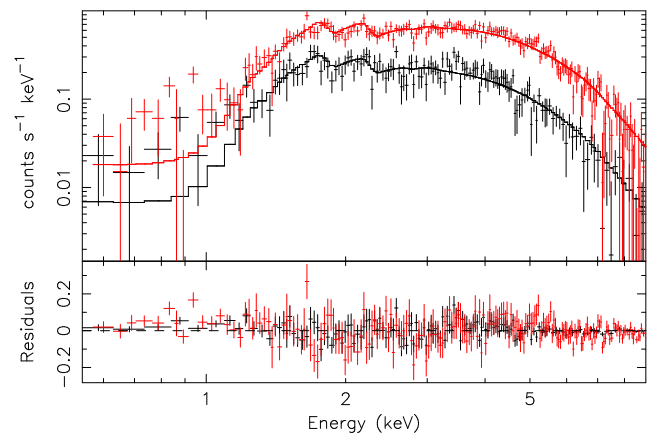}
  \caption[]{Time selected MOS1 count spectra from the 2003 September
  17 observation. The upper spectrum is from intervals when the
  0.5--10~keV count rate was $>$2.5~s$^{-1}$, the lower one from intervals
  below this value. The best-fit absorbed power-law models
  with the same value of $\alpha$ and different values of the
  normalisation and \nh\ for each spectrum are shown as histograms.}
  \label{fig:sep17_spec}
\end{figure}

\begin{table*}
\begin{center}
\caption[]{Summary of spectral fit parameters for the 2003 September observations.}
\begin{tabular}{l c c c c}
\hline
\hline\noalign{\smallskip}
Spectrum               &      \nh      &     $\alpha$     &  \chisq/d.o.f & 0.5-10 keV flux     \\
                    & (10$^{22}$ atom cm$^{-2}$)   &     &               & (erg cm$^{-2}$ s$^{-1}$)  \\ 
\hline\noalign{\smallskip}
Sep 11 (pre flare)  & 1.9$\pm$0.5   &		       &	       &     8.8$\times$10$^{-13}$    \\  
Sep 11 (flare)      & 4.3$\pm$0.2   & 2.20$\pm$0.09    &    151.6/151  &     4.0$\times$10$^{-11}$    \\ 
Sep 11 (post flare) & 3.1$\pm$0.3   & 		       &	       &     9.7$\times$10$^{-12}$    \\
                    &               &  		       &	       &                       \\   
Sep 11 (all)        & 3.8$\pm$0.4   & 2.25$\pm$0.15    &    151.9/142  &     1.5$\times$10$^{-11}$    \\
                    &               &		       &	       &                       \\
Sep 17 (high)       & 2.0$\pm$0.2   &\multirow{2}{*}{1.43$\pm$0.09}    & \multirow{2}{*}{430.3/347} &     
3.9$\times$10$^{-11}$    \\
Sep 17 (low)        & 2.7$\pm$0.2   &                  &               &     1.1$\times$10$^{-10}$    \\
                    &               & 		       &	       &                       \\  
Sep 17 (all)        & 2.5$\pm$0.2   & 1.42$\pm$0.09    &    283.8/215  &     7.5$\times$10$^{-11}$    \\
\noalign{\smallskip}
\hline
\label{tab:fit}
\end{tabular}
\end{center}
\end{table*}

Next, the spectral changes during the 2003 September 11 observation
were investigated. Three pn spectra were extracted corresponding to
the pre-flare (until 2003 September 11 20:02 UTC), flare (September~11
20:02 to 20:40~UTC), and post-flare (after September 11~20:40~UTC)
intervals. Background spectra were obtained simultaneously with the
source spectra, since significant variations in the background count
rate during the observation are present.  The absorbed power-law model
was used and the 3 spectra were fit simultaneously. First the value of
$\alpha$ was constrained to be the same for all 3 spectra and the
power-law normalisation, $k$, and $N _{\rm H}$ allowed to vary
independently. This gives an acceptable fit with a $\chi ^2$/d.o.f.\
of 151.6/151.  The best-fit value of $\alpha$ is $2.20 \pm 0.09$ and
\nh\ increases from $(1.9 \pm 0.5) \times 10^{22}$\,atom\,cm$^{-2}$ to
$(4.3 \pm 0.2) \times 10^{22}$\,atom\,cm$^{-2}$. The pre-flare
\nh\ is consistent with the interstellar value in the direction 
of \src\ of $1.4 \times 10^{22}$\,atom\,cm$^{-2}$ 
(Dickey \& Lockman~\cite{d:90}). However, it is also possible to
obtain a fit of similar quality if instead 
$\alpha$ is allowed to be different for the 3 spectra. Thus,
it is not possible to draw any firm conclusions on the nature
of the spectral changes {\it during} the observation, other than to say
that they are consistent with either a changing \nh\ {\it or} $\alpha$.   

In order to investigate whether similar variability was present during
the 2003 September 17 observation the MOS1 data were subdivided in two
intensity selected intervals above and below 2.5~count\,s$^{-1}$ and
spectra extracted.  MOS1 data were used due to the significantly
longer exposure compared to the pn. The 0.5--10\,keV intensities were
3.9$\times$10$^{-11}$\,erg\,cm$^{-2}$\,s$^{-1}$ and
1.1$\times$10$^{-10}$\,erg\,cm$^{-2}$\,s$^{-1}$.  The spectra were
again fit together allowing either $\alpha$ or \nh\ to vary (together
with the normalisation).  Again, it is not possible to draw any firm
conclusions since the variations are consistent with both a changing
\nh\ {\it or} $\alpha$. With $\alpha$ constrained to be the same, the
best-fit value of $\alpha$ is $1.43 \pm 0.09$ and \nh\ varies from
$(1.97 \pm 0.15) \times 10^{22}$\,atom\,cm$^{-2}$ to $(2.73 \pm 0.19)
\times 10^{22}$\,atom\,cm$^{-2}$ for a $\chi ^2$/d.o.f.\ of 430.3/347.
We note that the fit to the high-intensity spectrum is worse than to
the low-intensity spectrum. This may be caused by larger spectral
variations at high count rate. A summary of the fits to the different
spectra is given in Table \ref{tab:fit}.

\section{Discussion}
\label{sect:discussion}

The mean 0.5--10\,keV unabsorbed luminosity of \src\ during the 2003
September 11 and 17 observations is
1.1$\times$10$^{35}$\,erg\,s$^{-1}$ and
5.7$\times$10$^{35}$\,erg\,s$^{-1}$ for an (assumed) distance of
8\,kpc. The peak reached during the flare on 2003 September 17 is
about 8.5$\times$10$^{35}$\,erg\,s$^{-1}$ Such luminosities are only
reached in systems where the compact object is a neutron star or black
hole.  The state observed in 2003 March corresponds to an 0.5--10\,keV
luminosity of $<$4$\times$10$^{32}$\,erg\,s$^{-1}$. This upper limit
is consistent with the luminosities observed for quiescent X-ray
binary transients containing either a neutron star (e.g., Campana \&
Stella~\cite{cs:03}) or a black hole (e.g., Tomsick et
al.~\cite{t:03}).  The total dynamic flux range (quiescence to peak
flaring) seen during the three XMM-Newton observations is a factor
$\approxgt$2000.  This is also typical for such transient X-ray
binaries.

The (maximum) observed fluxes from the INTEGRAL observations (Sunyaev
et al.~\cite{s:03}) may constrain the spectral model in the soft and
hard X-ray bands, as well as the level of X-ray activity. Given the
quoted fluxes, and assuming that the hard X-ray spectrum consists of a
single power-law, the photon index was $\sim$4 during the INTEGRAL
observations. Assuming the interstellar \nh, this would give an
extrapolated absorbed 0.5--10\,keV flux of $\sim$6.4\,Crab. If during
the INTEGRAL observations the spectral index was $\sim$2.2, as
measured in the 0.5--10\,keV energy range on 2003 September 11, a
high-energy cut-off at $\sim$14\,keV is required in the spectrum in
order to explain the (maximum) 18--25\,keV and 25--60\,keV fluxes and
the 50--100\,keV flux upper limit. Extrapolating this spectrum gives
an absorbed flux of $\sim$0.45\, Crab (0.5--10\,keV).  Note that this
value would correspond to an unabsorbed luminosity of about
3$\times$10$^{38}$\,erg\,s$^{-1}$ (at 8\,kpc), similar to that reached
by classical X-ray binary transients (e.g., Chen et al.\ 1997).
Unfortunately, there are no closeby (i.e., within hours) {\it RXTE}
All-Sky Monitor (ASM) measurements of \src\ (R.~Remillard, private
communication), to verify whether the source was active around the
time of the INTEGRAL observations. The {\it RXTE}/ASM measurements
within days of the INTEGRAL detections give typical upper limits of
$\sim$50\,mCrab.  But since the source is highly variable in both soft
and hard X-rays, these do not provide stringent constraints either.

We note that the quiescent state and the low-level flaring seen with
XMM-Newton and the high-level activity seen by INTEGRAL is reminiscent
of SAX\,J1819.3$-$2525 (V4641\,Sgr).  This system also shows low-level
activity around 10$^{36}$\,erg\,s$^{-1}$ (in 't Zand et al.\ 2000),
strong and short high-level activity (e.g., Revnivtsev et
al.~\cite{re:02}), while in quiescence it reaches
$\simeq$4$\times$10$^{31}$\,erg\,s$^{-1}$ (0.3--8\,keV; Tomsick et
al.~\cite{t:03}).  The compact object in SAX\,J1819.3$-$2525 is most
probably a black hole (Orosz et al.~\cite{o:01}).

Our observed OM magnitudes, combined with the optical/infrared
magnitudes reported by Rodriguez et al. (\cite{r:03a}), and assuming
an absorption of 2$\times10^{22}$
\hcm\ and a distance of 8 kpc, are consistent with an early O-type companion.
However, a foreground object cannot be ruled out; we note the possible
presence of fainter optical candidates in the XMM-Newton error circle
(Rodriguez et al. \cite{r:03a}).

Future observations will hopefully shed more light on \src. In
particular, monitoring campaigns by INTEGRAL and multi-wavelength
observations may allow the nature of the compact object to be
elucidated.

\begin{acknowledgements}
%always
Based on observations obtained with XMM-Newton, an ESA science mission
with instruments and contributions directly funded by ESA member
states and the USA (NASA).

\end{acknowledgements}

\end{document}